# Investigation of Temperature Dependent Optical Modes in $Ge_xAs_{35-x}Se_{65}$ Thin Films: Structure Specific Raman, FIR and Optical Absorption Spectroscopy

Pritam Khan[1], Arinjoy Bhattacharya[1], Abin Joshy[1], Vasant Sathe[2], Uday Deshpande[2] and K. V. Adarsh[1]*

[1]*Department of Physics, Indian Institute of Science Education and Research, Bhopal 462023, India.*

[2]*UGC-DAE Consortium for Scientific Research, University Campus, Khandwa Road, Indore 452017,*

*India*

In this article, we present a comprehensive study of temperature and composition dependent Raman spectroscopy of $Ge_xAs_{35-x}Se_{65}$ thin films to understand different structural units responsible for optical properties. Strikingly, our experimental results uncover the ratio of $GeSe_{4/2}$ tetrahedral and $AsSe_{3/2}$ pyramidal units in $Ge_xAs_{35-x}Se_{65}$ thin films and their linear scaling relationship with temperature and x. An important notable outcome of our study is the formation of $Se_8$ rings at lower temperatures. Our experimental results further provide interesting optical features–thermally and compositionally tunable optical absorption spectra. Detailed structure specific FIR data at room temperature also present direct information on the structural units in consistent with Raman data. We foresee that our studies are useful in determining the lightinduced response of these films and also for their potential applications in optics and optoelectronics.

______________________________

*Author to whom correspondence should be addressed, electronic mail:  adarsh@iiserb.ac.in

**I. INTRODUCTION**

Ge-As-Se is the prototype of ternary amorphous chalcogenides (ChGs) and has been studied intensively in light of their technological applications in a diverse filed of optics and optoelectronics[1]: designing waveguides,[2] fabricating micro lens,[3] and recording holograms[4] etc. The main advantage of this system is the very broad glass-forming region that allows wide tuning of physical and optical properties and second, similarity in size and electronegativity of the constituent components, providing close-to-ideal covalent network. Consequently, it is very important to recognize the nature of amorphous networks to identify the different structural units forming the basic building blocks of the network. Such studies are therefore necessary to get detailed understanding on the underlying mechanism of several physico-chemical properties under various experimental conditions. For example, while illuminated with light[5], suspended in extreme pressure[6] or exposed to highly contrasting temperatures[7] etc.

In past, extensive studies (both Raman and FIR) have been devoted to identify the structural units of binary Ge-Se[8,9] or As-Se[10] glasses. On the otherhand, similar studies on that of ternary Ge-As-Se ChGs were scarcely depicted except for the work by Khan *et al.*[11] and Prasad *et al.*[12] because of extremely complicated network connection. According to their work, $GeSe_{4/2}$ tetrahedra and $AsSe_{3/2}$ pyramidal units are the basic building blocks in Ge-As-Se network. However, all these experiments have been performed in room temperature and hence temperature dependent response of these optical modes is completely missing. Nevertheless, it is required to address some of the important issues: (a) to understand the temperature dependent response of optical modes – $GeSe_{4/2}$ tetrahedra and $AsSe_{3/2}$ pyramidal units by Raman measurements and (b) to uncover the role of composition and temperature in determining the optical properties of $Ge_xAs_{35-x}Se_{65}$ network. Noteworthy to mention, all the structure specific measurements on ternary Ge-As-Se network glasses have been performed only with Raman spectroscopy and low frequency IR data (FIR) for such system is very insufficient. However it is essential to recognize IR data for complete assignment of vibrational bands to collect essential information regarding the thermodynamic properties. At times, Raman spectra is unable to provide low frequency information because of forbidden selection rule, weak Raman band, instability or color of the sample which makes them incompatible for experiments . To overcome these problems, the only plausible way is FIR measurements. In our present study, we have chosen ternary a-$Ge_xAs_{35-x}Se_{65}$ thin films. The main interest of dealing with this system is that it offers a parameter x (atomic percentage of germanium) which can vary continuously, leading to different measurable effects.[13] Furthermore the variation of x induces significant changes in optical properties related to the electronic structure near the absorption edge, as well as changes in the Raman and FIR spectrum related to the local atomic structure.

In this paper we present a detailed Raman spectroscopic study of a-$Ge_xAs_{35-x}Se_{65}$ thin films over a wide range of temperature starting from 90 to 390 K. Importantly, our results show that the ratio of $GeSe_{4/2}$ tetrahedral and $AsSe_{3/2}$ pyramidal unit scales a linear relationship with both temperature and x (Ge concentration). Further, at lower temperatures, we could observe the formation of $Se_8$ rings. Consequently, we found that optical absorption (OA) spectra of a-$Ge_xAs_{35-x}Se_{65}$ thin films is compositionally and thermally tunable–blue shifts (to shorter wavelength) with increase in x and redshift with increase in temperature. Subsequent structure specific FIR measurements at room temperatures also provide valuable information on the various optical modes in consistent with Raman data.

**II. Experimental**

**A. Film preparation**.

Bulk samples of $Ge_xAs_{35-x}Se_{65}$ were prepared from stoichiometric mixture of high pure (99.999%) Ge, As and Se powders. The purified raw materials were weighed with the precision of 0.01% and kept in a quartz ampoule which was evacuated vacuum of ~$1\times10^{-6}$ mbar and sealed. The sealed ampoule was kept inside a furnace of 800°C and rotated continuously for about 24 hours to ensure homogeneity of the melt. Bulk glass was obtained by quenching the ampoule in ice cooled water. $Ge_xAs_{35-x}Se_{65}$ thin films of ~1 μm thickness were deposited on a microscope glass substrate by conventional thermal evaporation technique in a vacuum of $5\times10^{-6}$ mbar. The deposition rate was kept below 5A°/s was maintained and continuously monitored using a quartz crystal. It is well known that such a low deposition rate produces thin films of composition which is very close to that of starting bulk samples. We have later performed EDAX and confirmed the uniformity of composition within the experimental error of this technique, i.e. within 2-3 %.

**B. Structural characterization.**

Raman spectra were collected in backscattering geometry using the 632 nm excitation light from a He-Ne laser of power 4.5 mW coupled with a Labram-HR800 micro-Raman spectrometer equipped with a X 50 objective, an appropriate notch filter, and a Peltier cooled charge coupled device detector. For the low temperature Raman measurements, the sample was mounted on a THMS600 stage from Linkam UK, with temperature stability of ± 0.1 K. It should be noted that spectral resolution of the Raman system is ~1cm$^{-1}$.

Transmission spectra of the sample were meaured using a high resolution Ocean Optics HR 4000 spectrometer. Temperature dependent spectra was recoreded by putting the sample inside an optical cryostat having a temperature scan range 80 to 400 K.

Far infrared (FIR) absorption spectra of all samples were measured at room temperature by using Bruker Vertex 70 spectrometer. The wavelength of the probe beam was in the range of 100 to 400 cm$^{-1}$. For FIR measurements, thin films were coated on the polyethylene substrate which shows good transparency in the FIR region.

**III. Results and Discussions**

**A. Room temperature Raman spectroscopy of a-$Ge_xAs_{35-x}Se_{65}$ thin films**

To get detailed understanding on the different structural units of $Ge_xAs_{35-x}Se_{65}$ thin films, we have recorded the Raman spectra of all films as shown in figure 1(a). Importantly, dominant features of Raman spectra in all the samples are in the range of approximately 190 to 270 cm$^{-1}$. Such observation indicates the similarity in mass and bond strengths among the three constituent components. They mainly consist of two independent modes: (1) a sharp peak at 198 cm$^{-1}$ and (2) a broad peak that extends from 220–245 cm$^{-1}$. The Raman peak at 198 cm$^{-1}$ is assigned to $A_1$ ($v_1$) symmetric vibrational stretching of $GeSe_{4/2}$ corner-sharing tetrahedra (CST).[6,9] Importantly, the magnitude of this peak increases with increase in Ge concentration – the number of defect bonds increases resulting in an increase in the strength of the defect induced absorption tail within the bandgap.[12] The broad peak from 220–245 cm$^{-1}$ is attributed to the principal vibrational modes of $AsSe_{3/2}$ pyramidal (ASP) unit and also to minor contributions from $A_1$ ($v_2$) modes of $As_4Se_3$ cage like molecule.[6,9,14] Notably, these bands are asymmetric in nature which might be arising from the contribution of As–As–Se$_2$ structure and shifts to longer wavenumber when Ge concentration increases from 5 to 25. In addition to these two dominant features, samples with x=15 and 25 show a peak at 215 cm$^{-1}$, identified as the companion mode originating from the vibrational edge sharing GeSe$_{4/2}$ tetrahedra (EST) [6,9] which lies in the breathing vibrations of distorted fragments of layered c-GeSe$_2$. Importantly, this mode is present only in samples with high Ge concentration, precisely when x ≥15 in our present series.

Since Ge-Se and As-Se are the basic building block units of Ge-As-Se network glasses, it is important to understand quantitatively the individual contributions from vibrational modes of Ge-Se and As-Se units. In this context, we have plotted in figure 1(b) peak intensity ratio of 198 cm$^{-1}$ due to GeSe$_{4/2}$ CST and to that of ASP unit at 220–245 cm$^{-1}$ with x. Interestingly, we found that the experimental data fit linearly which unveils the linear scaling relationship of this ratio with Ge concentration in the glasses.

**B. Temperature dependent optical modes in a-Ge$_x$As$_{35-x}$Se$_{65}$ thin films: Raman Spectroscopic study**

Next, to understand qualitatively the temperature dependent response of optical modes of a-Ge$_x$As$_{35-x}$Se$_{65}$ thin films, we have performed Raman spectroscopic measurements of all samples over a wide temperature range. In this context, figure 2(a)-(d) represents the Raman spectra of a-Ge$_x$As$_{35-x}$Se$_{65}$ thin films at three contrasting temperatures: 90, 300 and 390 K. Interestingly, we could observe an appreciable enhancement in Raman signal for corner sharing GeSe$_{4/2}$ tetrahedra at 198 cm$^{-1}$ ( more significantly for x≥15), when the temperature is raised from 90 to 390 K. However, peak position remains unaltered with respect to temperature. On the otherhand, for AsSe$_{3/2}$ pyramidal (ASP) unit spreading across 220–245 cm$^{-1}$, the peak position is same at room and high temperatures. Nonetheless at 90 K, this band shifts to lower wavenumber displaying a broad band centred at ~ 250 cm$^{-1}$. It is believed that this broad peak is associated with A$_1$ modes of Se$_8$ rings.[9,13,15] Our results therefore provide first evidence of temperature mediated structural rearrangements in a-Ge$_x$As$_{35-x}$Se$_{65}$ thin films.

As already mentioned, Ge-As-Se network is primarily based on Ge-Se and As-Se building blocks so it is important to quantify the relative contribution of these units at different temperatures. In this regard, we have analyzed the Raman data of all the samples in detail and calculated the peak intensity ratio of the Raman mode corresponding to corner sharing GeSe$_{4/2}$ tetrahedra (CST) and AsSe$_{3/2}$ pyramidal (ASP) units at all measuring temperatures as shown in figure 3(a)-(d). The results show that with increasing temperature the ratio of GeSe$_{4/2}$ CST and ASP increases – GeSe$_{4/2}$ CST dominates over ASP at higher temperatures and will play the predominant role in determining various physical properties. Evidently, a comparative study between all samples indicate that the relative change in the ratio of GeSe$_{4/2}$ CST to ASP is maximum for x=25 and minimum for that of x=5. Likewise, at any measuring temperature, the ratio of Ge-Se/As-Se bond density is largest for Ge rich sample (Ge$_{25}$As$_{10}$Se$_{65}$) when compared to other Ge deficient samples.

**C. Compositional analysis by Optical Absorption (OA) Spectroscopy**

Figure 4(a) shows the OA spectra of a-Ge$_x$As$_{35-x}$Se$_{65}$ thin films at room temperature. It is quite clear from the figure that absorption spectra blue shift when x is increased from 5 to 25, in other words from Ge deficient to Ge rich samples. At this point, we believe that the blue shift in OA spectra of a-Ge$_x$As$_{35-x}$Se$_{65}$ thin films might be related to the shift of AsSe$_{3/2}$ pyramidal (ASP) unit to longer wavenumber with increase in x (Ge concentration) as evident from Raman spectra (figure 2(a)). This indicates that OA spectra are compositionally tunable. At this point, it is worthy to mention that similar blueshift has also been observed by Kumar *et al.* for binary Ge$_x$Se$_{100-x}$ system.[16]

Next, we have calculated the optical bandgap of each sample by using Tauc equation[17]

$$(\alpha h\nu)^{1/2} = B^{1/2}(h\nu - E_g) \qquad (1)$$

where α, h, ν, E$_g$ and B are the absorption coefficient, Plank′s constant, frequency, optical band gap, and a constant (Tauc parameter) respectively. A straight like fit to the plot (αhν)$^{1/2}$ vs hν gives the value of optical bandgap.[18] To understand quantitatively, we have plotted in figure 4(b) the variation of bandgap with Ge composition. Evidently, bandgap increases with increasing x, i.e. when we move from Ge deficient to Ge rich samples.

## D. Temperature dependence of OA spectra: Variation of optical bandgap as a function of temperature.

To uncover the role of temperature on the bandgap of the sample, we have recorded the OA spectra of a-$Ge_xAs_{35-x}Se_{65}$ thin films over a wide temperature range and are shown in figure 6(a)-(d). Interestingly, we could observe a red shift in the OA spectra of all samples when the temperature is raised from 90 to 390 K. This redshift can be explained as a consequence of the change in degree of deviation from the periodicity of the lattice vibrations with temperature.[19] Consequently, optical bandgap changes with temperature. To substantiate this idea, we have plotted in figure 6 the optical bandgap of individual sample as a function of measuring temperatures. It is well apparent from the figure that for all samples bandgap decreases with increasing temperature. Such observation can be explained by the temperature dependent electron-phonon interaction which effectively determines semiconductor bandgap.[20] Precisely, bandgap reflects bond energy and an increase in temperature changes the chemical bonding as electrons are promoted from valence band to conduction band which reduces the optical bandgap at higher temperatures. Nevertheless, at any temperature, bandgap of Ge rich sample is always more than the Ge deficient samples.

## E. Structure specific Far Infra Red (FIR) absorption spectroscopy in a-$Ge_xAs_{35-x}Se_{65}$ thin films.

To get new insights on different structural units of a-$Ge_xAs_{35-x}Se_{65}$ thin films we have performed FIR absorption spectroscopy of all samples at room temperature. To ensure that the substrate has no contribution to the observed peaks of our samples, we have recorded the absorption spectra of the blank substrate and made that as background. In this regard, figure 7(a) shows the FIR absorption spectra of four samples in the frequency range of 185-325 $cm^{-1}$. The FIR absorption spectra of all samples consist of two independent modes: (1) a broad peak ($M_1$) spreads from 220-240 $cm^{-1}$, attributed to the $v_7$ modes of As-Se structural unit and a contribution from $A_1$ and E modes of Se polymeric chain[15,21] and (2) a relatively sharper peak ($M_2$) centred at 256 $cm^{-1}$ which is assigned to the asymmetric bond-stretching mode $F_2$ of $GeSe_{4/2}$ tetrahedra[22,23] and a weak contribution from $Se_8$ ring ($A_1$, E mode).[24,25] A quick comparison between Raman and FIR spectra reveals that the dominant peak of 198 $cm^{-1}$ which appears for all samples in Raman spectra is absent for FIR spectra. It is quite natural because as already mentioned the peak at 198 $cm^{-1}$ is assigned to corner sharing $GeSe_{4/2}$ tetrahedra with symmetric breathing of Se atoms which cannot change the dipole moment of vibrating unit, therefore remains inactive in FIR spectra.[26] From the figure 7(a) it is apparent that $M_1$ mode shifts to higher wavenumber side (221 to 239 cm$^{-1}$) with increasing Ge concentration whereas $M_2$ is resistant against composition variation and centred at 256 cm$^{-1}$ for all samples. Moreover, magnitude of both $M_1$ and $M_2$ is found to vary with composition. Clearly those two modes exhibit contrasting characteristics, i.e. magnitude of $M_1$ decreases whereas that for $M_2$ increases when x increases from 5 to 25. It is expected because the mode $M_2$ is assigned to GeSe$_{4/2}$ tetrahedra due to Ge-Se bonds and likewise it will increase with increasing Ge concentration. At this point, we will try to explain the decrease in optical bandgap with decrease in Ge concentration (figure 1(b)), in other words with increasing As concentration. Importantly, as the As concentration increases heteropolar As-Se bond starts forming on the expense of Se-Se bonds which effectively reduces the average energy of the system. This inturn results in the decrease of optical bandgap with increasing As concentration. To understand quantitatively, we have analyzed the FIR data in more detail and calculated the absorption peak ratio of different vibrational modes $M_1$ and $M_2$. As can be seen from figure 8 (b) ratio of $M_2/M_1$ increases with increasing Ge concentration, i.e. $F_2$ of GeSe$_{4/2}$ tetrahedra dominates over $\nu_7$ modes of As-Se structural unit as we move from Ge deficient to Ge rich samples which is quite similar to what we observed from Raman spectra.

The utmost purpose of performing structure specific Raman, FIR and optical absorption spectra of a-Ge$_x$As$_{35-x}$Se$_{65}$ thin films over a wide range of temperature are to identify the different optical modes and to study their response at extreme contrasting temperatures. Such studies are very important to understand the underlying mechanism of various physico-chemical properties giving rise to numerous photoinduced phenomena in ChGs, e.g. photodarkening, photobleaching, optical anisotropy etc. A detailed understanding with new insights on structural units is indeed necessary for synthesizing photostable / photo-sensitive glasses for various applications owing to their stability/changes in optical properties induced by annealing and light exposure.

**IV. Conclusions**

Our systematic Raman and FIR studies in a-Ge$_x$As$_{35-x}$Se$_{65}$ thin films give new insight on the GeSe$_{4/2}$ tetrahedra and AsSe$_{3/2}$ pyramidal units that form the amorphous network and their linear scaling relationship with x and temperature. A notable observation of our study is that formation of Se$_8$ rings at lower temperatures which give the first evidence of temperature dependent structural rearrangements. Interestingly, OA blue shifts with increasing x in consistent with the shift observed for $AsSe_{3/2}$ pyramidal (ASP) unit towards longer wavenumber. Needless to say, our results demonstrate temperature and composition dependent structural rearrangements in a-$Ge_xAs_{35-x}Se_{65}$ thin films. We foresee that our studies are helpful in understanding the physico-chemical properties and lightinduced effects of these thin films which can hopefully contribute to the field of optics and optoelectronics.

**Acknowledgements**

The authors thank Department of Science and Technology (Project no: SR/S2/LOP-003/2010) and Council of Scientific and Industrial Research, India, (grant No. 03(1250)/12/EMR-II) for financial support.

**References**

[5]B. Vaidhyanathan, S. Murugavel, S. Asokan, and K. J. Rao, J. Phys. Chem. B **101**, 9717 (1997)

[1]K. Shimakawa, A. Kolobov, and S. R. Elliott, Adv. Phys. **44**, 475 (1995).

[2]S. Spälter, H. Y. Hwang, J. Zimmermann, G. Lenz, T. Katsufuji, S. –W. Cheong, and R. E. Slusher, Opt. Lett. **27**, 363 (2002).

[3]H. Hisakuni, and K. Tanaka, Opt. Express **20**, 958 (1995).

[4]Y. Somemura, A. Yoshikawa, and Y. Utsugi, Jpn. J. Appl. Phys. **31**, 3712 (1992).

[5]P. Khan, A. R. Barik, E. M. Vinod, K. S. Sangunni, H. Jain, and K. V. Adarsh, Opt. Express **20,**12416 (2012).

[6]S. Asokan, M. V. N. Prasad, G. Parthasarathy, and E. S. R. Gopal, Phys. Rev. B **62**, 808 (1989).

[7]P. Khan, H. Jain, and K. V. Adarsh, Sci. rep. **4,** 4029 (2014).

[8]S. R. Elliott, Nature **354**, 445 (1991).

[9]S. Sugai, Phys. Rev. B **35**, 1345 (1987).

[10]V. Kovandaa, M. Vlček, and H. Jain, J. Non-Cryst. Solids **326**, 88 (2003).

[11]P. Khan, T. Saxena, H. Jain, and K. V. Adarsh, Sci. Rep. **4**, 6573 (2014).

[12]A. Prasad, A.; C. –J. Zha, R. P. Wang, A. Smith, S. Madden, and B. L. Davies, Opt. Express **16**, 2804 (2008).

[13]P. Tronc, M. Bensoussan, A. Brenac, and C. Sebenne, Phys. Rev. B **8**, 5947 (1973).

[14]X. Su, R. Wang, B. L. Davies, and L. Wang, Appl. Phys. A **113**, 575 (2013).

[15]A. R. Barik, M. Bapna, R. Naik, U. Deshpande, T. Sripathi, and K. V. Adarsh, Mater. Chem. Phys. **138**, 479 (2012).

[16]R. R. Kumar, A. R. Barik, E. M. Vinod, M. Bapna, K. S. Sangunni, and K. V. Adarsh, Opt. Lett. **38**, 1682 (2013).

[17]J. Tauc, R. Grigorovici, and A. Vancu, Phys. Stat. Sol. **15**, 627 (1966).

[18]R. Naik, and R. Ganesan, J. Non-Cryst. Solids **385**, 142 (2014).

[19]H. Y. Fan, Phys. Rev. **82**, 900 (1951).

[20]K. P. O'Donnell, and X. Chen, Appl. Phys. Lett. **58**, 2924 (1991).

[21]G. J. Ball, and J. M. Chamberlain, J. Non-Cryst. Solids **29**, 239 (1978).

[22]Y. Utsugi, and Y. Mijushima, J. Appl. Phys. **49**, 3470 (1978).

[23]H. J. Trodahl, Solid State Commun. **44**, 319 (1982).

[24]P. Shamra, V. S. Rangra, P. Sharma, and S. C. Katyal, J. Alloys Compd. **480**, 934 (2009).

[25]G. Lucovsky, A. Mooradian, W. Taylor, G. B. Wright, R. C. Keezer, Solid State Commun. **5**, 113 (1967).

[26]P. Nemec, B. Frumarvo, and M. Frumar, J. Non-Cryst. Solids **270**, 137 (2000).

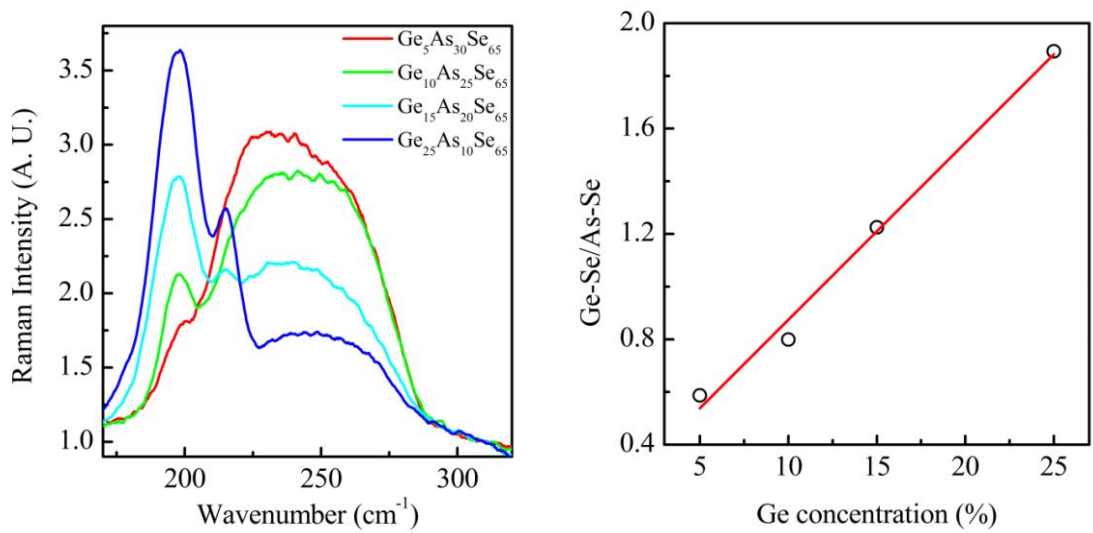

**Fig.1** (a) Raman spectra of a-Ge$_x$As$_{35-x}$Se$_{65}$ thin films. Clearly the spectra consists of (1) a sharp peak at 198 cm$^{-1}$ assigned to the A$_1$ (ν$_1$) symmetric vibrational stretching of GeSe$_{4/2}$ corner-sharing tetrahedra, (2) The broad peak from 220–245 cm$^{-1}$ is attributed to the principal vibrational modes of AsSe$_{3/2}$ pyramidal (ASP) unit and (3) a peak at 215 cm$^{-1}$ assigned vibrational edge sharing GeSe$_{4/2}$ tetrahedra which exists only for x ≥15. (b) Peak intensity ratio of 198 cm$^{-1}$ to 220–245 cm$^{-1}$ which scales linearly with Ge concentration. The black hollow circles and red line represents the experimental data and theoretical fit respectively.

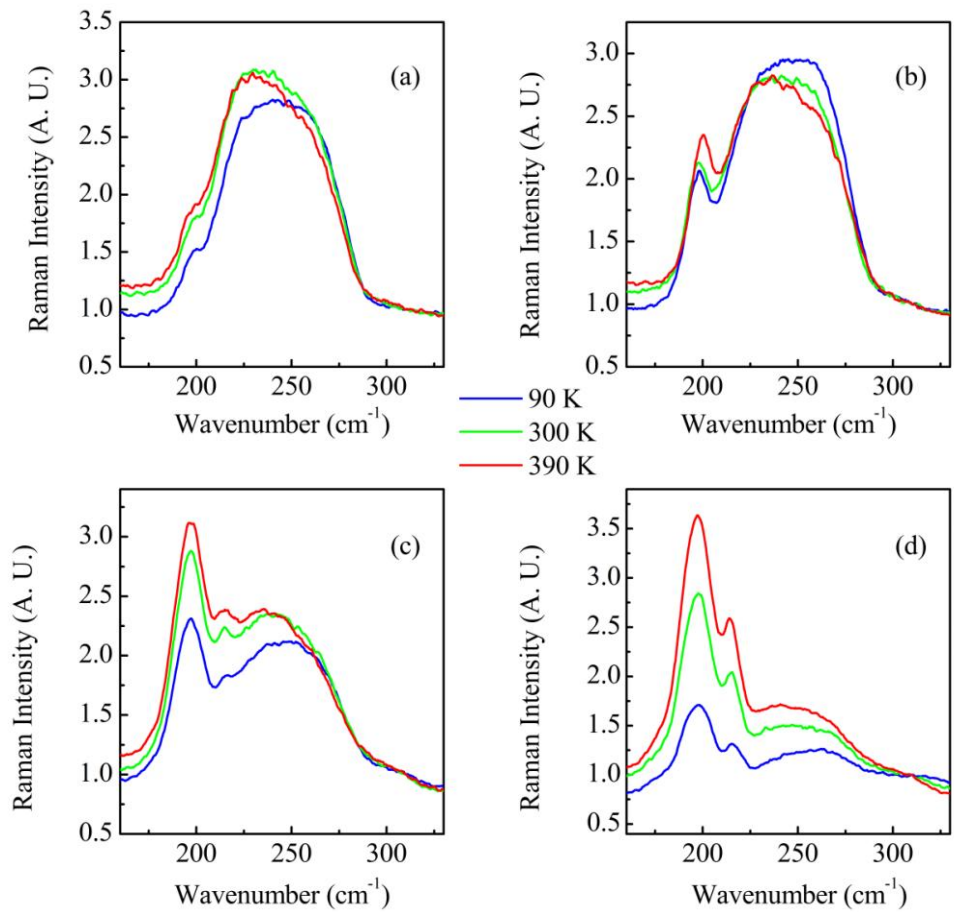

**Fig. 2** Raman spectra of (a) a-$Ge_5As_{30}Se_{65}$ (b) a-$Ge_{10}As_{25}Se_{65}$ (c) a-$Ge_{15}As_{20}Se_{65}$ and (d) a-$Ge_{25}As_{10}Se_{65}$ thin films at three contrasting temperatures 90, 300 and 390 K. Apparently, Raman signal for corner sharing $GeSe_{4/2}$ tetrahedra at 198 cm$^{-1}$ enhances with increasing temperature, whereas at lower temperatures $AsSe_{3/2}$ pyramidal (ASP) unit shifts toward higher wavenumber side being converted into $Se_8$ rings.

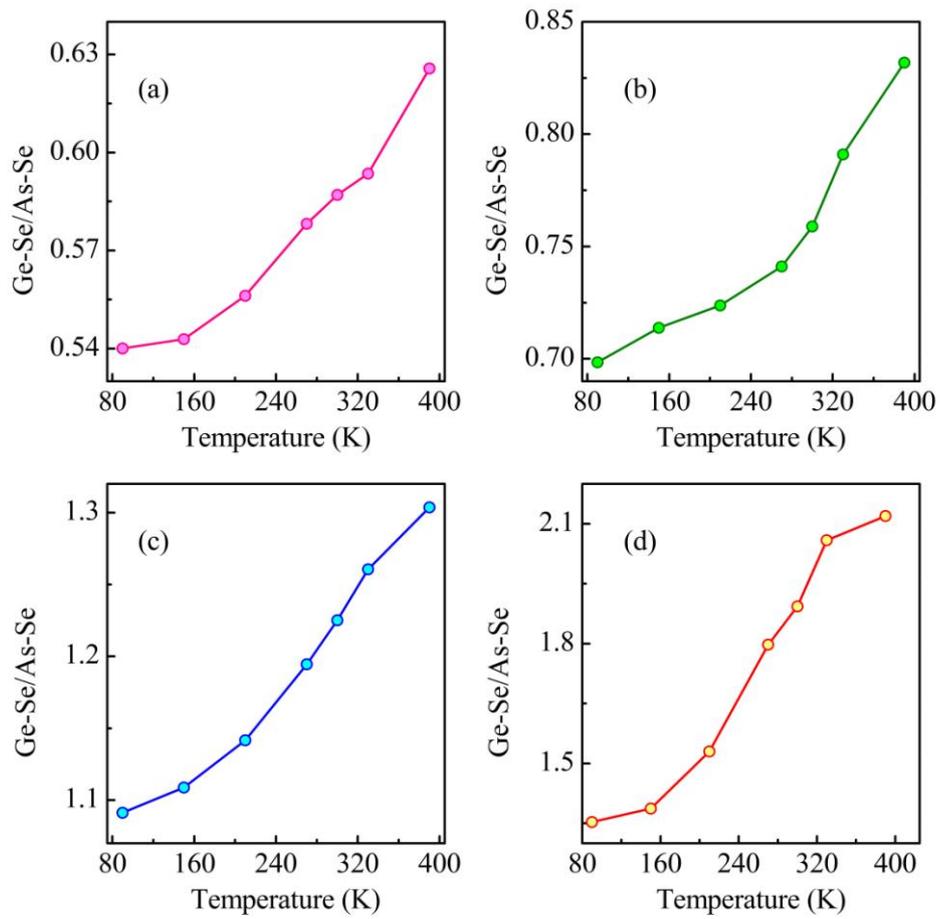

**Fig. 3** Peak intensity ratio of GeSe$_{4/2}$ tetrahedra to AsSe$_{3/2}$ pyramidal unit of (a) a-Ge$_5$As$_{30}$Se$_{65}$ (b) a-Ge$_{10}$As$_{25}$Se$_{65}$ (c) a-Ge$_{15}$As$_{20}$Se$_{65}$ and (d) a-Ge$_{25}$As$_{10}$Se$_{65}$ thin films for various measuring temperatures. Evidently, the ratio increases with increasing temperature signifying that GeSe$_{4/2}$ tetrahedra unit dominates over AsSe$_{3/2}$ pyramidal unit at higher temperatures.

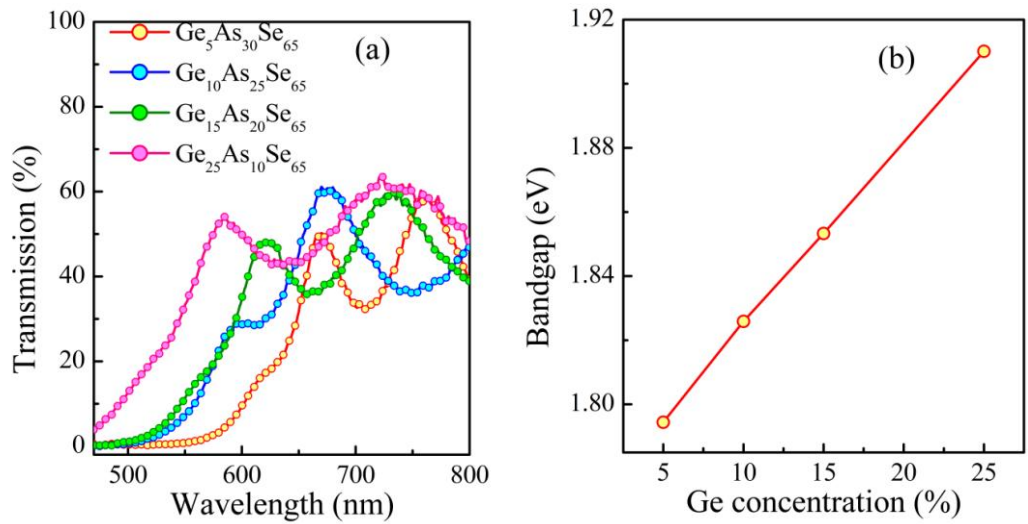

**Fig. 4** (a) Optical absorption spectra of a-Ge$_x$As$_{35-x}$Se$_{65}$ thin films. Interestingly, the spectra blue shifts when we move from Ge deficient to Ge rich samples. (b) Optical bandgap variation of a-Ge$_x$As$_{35-x}$Se$_{65}$ thin films which increases with increasing Ge concentration.

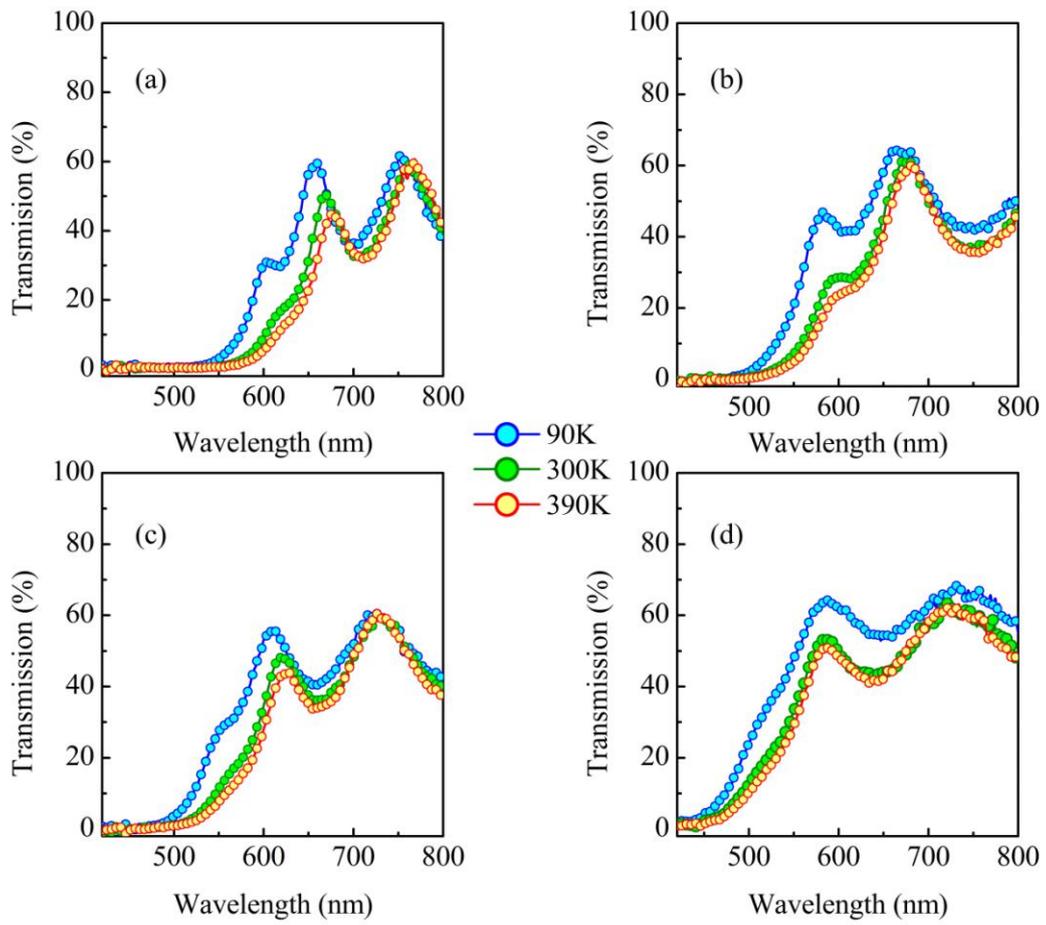

**Fig. 5** Optical absorption spectra of (a) a-Ge$_5$As$_{30}$Se$_{65}$ (b) a-Ge$_{10}$As$_{25}$Se$_{65}$ (c) a-Ge$_{15}$As$_{20}$Se$_{65}$ and (d) a-Ge$_{25}$As$_{10}$Se$_{65}$ thin films at 90, 300 and 390 K. Interestingly, for each sample OA spectra red shifts with increase in temperature.

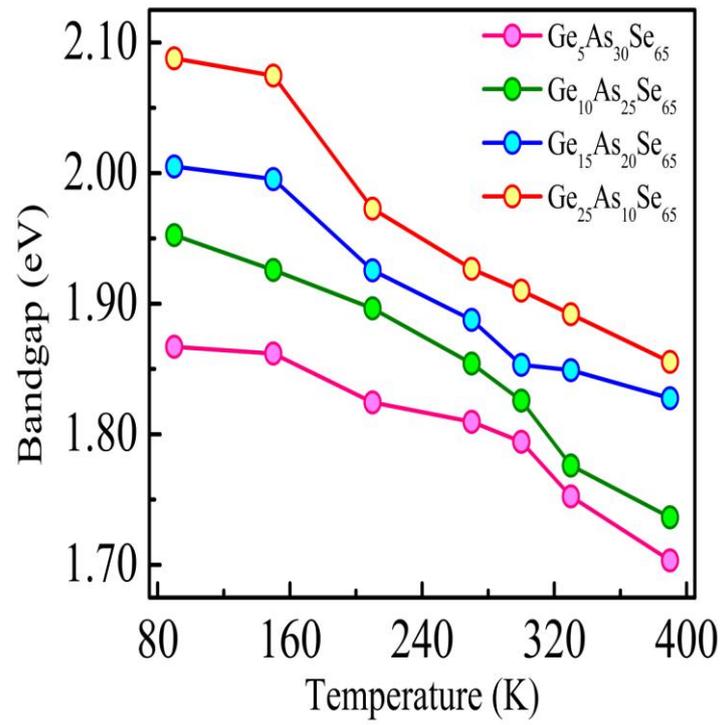

**Fig. 6** Temperature dependent optical bandgap variations of a-$Ge_xAs_{35-x}Se_{65}$ thin films which indicate that for all samples bandgap decreases with increase in temperature.

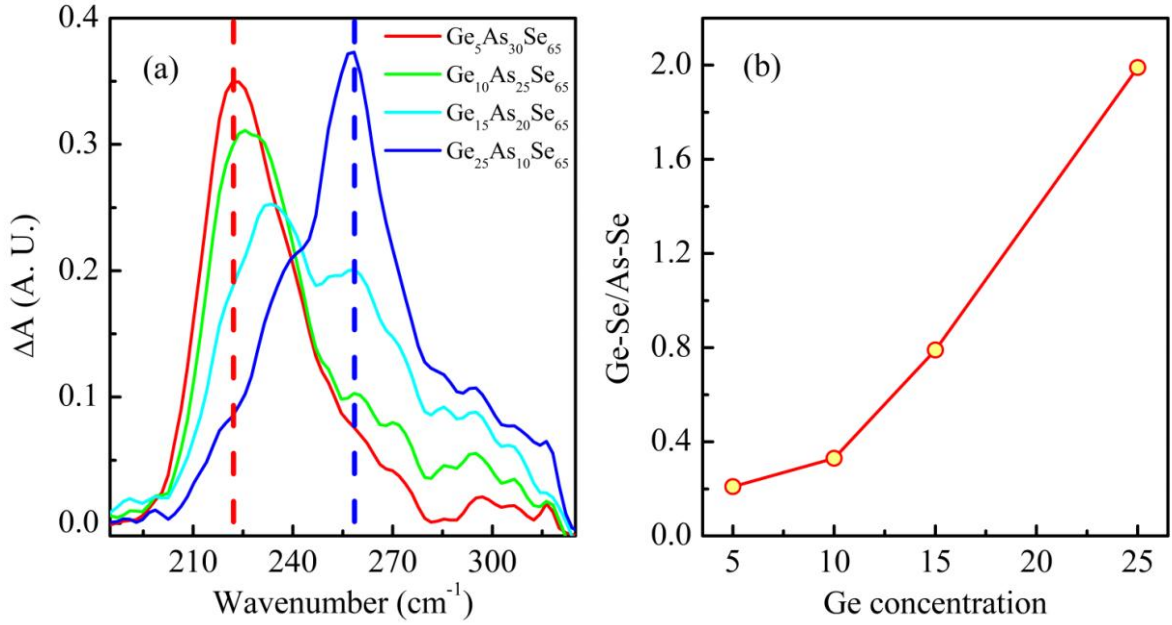

**Fig. 7** (a) FIR absorption spectra of a-$Ge_xAs_{35-x}Se_{65}$ thin films which consists of a broad peak ($M_1$) spreads from 220-240 cm$^{-1}$ (dotted red line), attributed to the $\nu_7$ modes of As-Se structural unit with a minor contribution from $A_1$ and E modes of Se polymeric chain and sharper peak ($M_2$) centred at 256 cm$^{-1}$ (dotted blue line), assigned to the asymmetric bond-stretching mode $F_2$ of $GeSe_{4/2}$ tetrahedra and minor contribution from $A_1$ and E modes of $Se_8$ ring. (b) Peak intensity ratio of $M_2$ to $M_1$ as a function of Ge concentration that increases as we move from Ge deficient to Ge rich samples.